\documentclass[a4paper]{jpconf}
\usepackage{graphicx}
\usepackage{iopams}

\newcommand{\kb}{k_\mathrm{B}}
\newcommand{\FDT}{{\small{\textsf{FDT}}}\,}
\newcommand{\FDR}{{\small{\textsf{FDR}}}\,}
\newcommand{\ave}[1]{\left\langle #1\right\rangle}

\begin{document}

\title{Effective temperatures in a simple model of non-equilibrium, non-Markovian dynamics}

\author{Patrick Ilg and Jean-Louis Barrat}

\address{Universit\'e Lyon1, Laboratoire de Physique de la Mati\`ere Condens\'ee
et Nanostructures,   F-69622 Villeurbanne Cedex, France; CNRS,
UMR5586}

\ead{Patrick.Ilg@lpmcn.univ-lyon1.fr}

\begin{abstract}
The concept of effective temperatures in nonequilibrium systems is
studied within an exactly solvable model of non-Markovian
diffusion. The system is coupled to two heat baths which are kept
at different temperatures: one ('fast') bath associated with an
uncorrelated Gaussian noise and a second ('slow') bath with an
exponential memory kernel. Various definitions of effective
temperatures proposed in the literature are evaluated and
compared. The range of validity of these definitions is discussed.
It is shown in particular, that the effective temperature defined
from the fluctuation-dissipation relation mirrors the temperature
of the slow bath in parameter regions corresponding to a
separation of time scales. On the contrary, quasi-static and
thermodynamic definitions of an effective temperature are found to
display the temperature of the fast bath in most parameter
regions.
\end{abstract}

\section{Introduction}

Recent years have witnessed enormous efforts in understanding
glassy systems by studying a broad variety of model systems.
At the same time, it has been found that various other
systems like colloids, granular materials, etc. share several features of
typical phenomenology of glassy systems.
Therefore, a common macroscopic description of glassy systems would
be extremely helpful.
However, unlike equilibrium thermodynamics, there is no general,
well-established framework for a macroscopic description available
today that works out of equilibrium.

A key role in the macroscopic description of glassy systems could
be played by the concept of effective temperature
\cite{Cugliandolo_effTemp97,Nieuwenhuizen_efftemp}. Based on the
fluctuation-dissipation relation (\FDR)
\cite{Cugliandolo_effTemp97}, the effective temperature is
supposed to govern the slow, not equilibrated degrees of freedom,
while the short time dynamics and the equilibrated degrees of
freedom are governed by the bath temperature. Subsequent molecular
dynamics simulations \cite{KobBarrat2000,Barrat_FDT2002} on a
supercooled model fluid during aging or in shear flow are in
agreement with this picture.
However, it has been argued that non-linear generalizations of the
fluctuation-dissipation theorem (\FDT) out of equilibrium could
spoil the interpretation of effective temperatures from \FDR
\cite{Jou_Review_nonequiTemp2003}. The authors emphasize that
``Physical interpretations of the several temperatures and their
relation to experimental probes and to their mutual relation in
non-equilibrium steady states are needed.'' (p.~2014 in
\cite{Jou_Review_nonequiTemp2003}). In addition, more recent
molecular dynamics simulations \cite{OHern2004} found that
effective temperatures should sometimes be defined from static
\FDR rather than from the dynamical one, as was done in
\cite{KobBarrat2000,Barrat_FDT2002}. The authors of
\cite{OHern2004} conclude their study with a ``puzzle: when should
one use static linear response and when should one use a
time-dependent relation?''. For a review on the numerical evidence
of effective temperatures see also
\cite{CristantiRitort_efftempreview}. Up to now, experimental
investigations of effective temperatures in glassy systems are not
conclusive. Experiments on spin glasses in the aging regime
\cite{Herrison2002} and self-diffusion of granular material in a
Couette cell \cite{Makse2005} have shown some evidence in favor of
the existence of an effective temperature as introduced in
\cite{Cugliandolo_effTemp97}. In dense colloidal suspensions, very
recent experiments \cite{Bonn2005} found no deviation from the
\FDT, while an effective temperature could be measured in a
colloidal glass of laponite \cite{Strachan2005,AbouGallet2004}.
Most experiments, however, involve frequencies that are rather
high compared to the inverse relaxation time.  Earlier experiments
on an aging colloidal glass \cite{Bellon2002} observed violations
of the \FDT in electrical but not in rheological studies.

In view of these controversies, it seems appropriate to resort
to simplified model systems, where the validity of
effective temperature concepts can be studied in more detail.
It has been found, that mean-field spin glasses with one-step
replica symmetry breaking scheme do allow the definition of an effective
temperature from the \FDR \cite{Cugliandolo_PRL93},
while those with continuous replica symmetry breaking do not
\cite{CristantiRitort_efftempreview}.
Field-theoretic calculations on the critical dynamics of spin models
show, that the effective temperature defined from the \FDR
does depend on the variables considered \cite{Gambassi,Schehr}.
Also in the Ising chain with Glauber dynamics and dynamical trap models,
a meaningful effective temperature cannot be defined from the \FDR
\cite{Sollich_condmat01}.
A slightly different perspective was taken in
\cite{Cugliandolo_FluktTheorem,Cugliandolo_scenario99,Nieuwenhuizen_steadystatedifftimescales},
where Brownian particles coupled to two heat baths at different
temperatures were studied.
The aim of the present study is to derive and compare different
static as well as dynamic
definitions of effective temperatures in these systems and
discuss their range of validity.

This paper organized as follows.
The model of non-Markovian diffusion in the presence of two heat
baths is introduced in Sec.~\ref{nonmarkov.sec}.
In Sec.~\ref{overdamped.sec}, the model is simplified by considering
the overdamped limit.
From the exact solution of the model,
different definitions of effective temperatures are evaluated and
compared in Sec.~\ref{harmonic.sec}.
Some conclusions are offered in Sec.~\ref{concl.sec}.

\section{Model} \label{nonmarkov.sec}
Consider a particle of mass $m$ at position $x$ with velocity $v$
moving in a potential $V(x)$ under the influence of two thermal baths.
One bath is held at temperature $T_\mathrm{slow}$.
Its influence on the dynamics
of the particle is described by the retarded friction coefficient
(memory kernel) $\Gamma(t)$.
The other bath is kept at temperature $T_\mathrm{fast}$.
Contrary to the first, slow bath, the correlation time of the second bath
is small enough, so that the particle experiences an instantaneous
friction described by the friction coefficient $\Gamma_0$.
The equations of motion read
\begin{eqnarray} \label{original}
\dot{x} & = & v \nonumber\\
m\dot{v} & = & - \frac{\partial V}{\partial x} - \int_0^t\!ds\,
\Gamma(t-s)v(s) -\Gamma_0v(t) + \xi(t) + \eta(t).
\end{eqnarray}
The fast bath is modelled as Gaussian white noise with
$\ave{\eta(t)}=0$ and
$\ave{\eta(t)\eta(s)}=2T_\mathrm{fast}\Gamma_0\delta(t-s)$,
whereas the random force due to the slow bath is characterized by
$\ave{\xi(t)\xi(s)}=T_\mathrm{slow} \Gamma(t-s)$. We choose units
such that Boltzmann's constant $\kb\equiv 1$. The model
(\ref{original}) generalizes earlier work on non-Markovian
diffusion \cite{Barrat_nonmarkov90} by including a second heat
bath at a different temperature.

The diffusion equation (\ref{original}) is difficult to study
in general due to its non-Markovian character.
For the special case of an exponential memory function,
\begin{equation} \label{memory}
  \Gamma(t) = \frac{1}{\alpha}\exp{[-t/(\alpha\gamma)]},
\end{equation}
it is possible to map the non-Markovian dynamics (\ref{original}) onto
a Markovian process in an extended state space \cite{Barrat_nonmarkov90},
\begin{eqnarray} \label{langevin}
\dot{x} & = & v \nonumber\\
m\dot{v} & = & - \frac{\partial V}{\partial x} + z(t)- \Gamma_0v(t) + \eta(t) \nonumber\\
\dot{z} & = & -\frac{1}{\alpha}v(t) - \frac{1}{\alpha\gamma}z(t) + \zeta(t)
\end{eqnarray}
where $\zeta$ is a Gaussian white noise with
\begin{equation} \label{R1R1}
  \ave{\zeta(t)\zeta(s)}=\frac{2T_\mathrm{slow}}{\alpha^2\gamma}\delta(t-s).
\end{equation}
Equations (\ref{langevin}) together with the noise correlators specify
the model to be considered in the present study.

\section{The overdamped limit} \label{overdamped.sec}
Formally, the model of non-Markovian diffusion introduced in
Sec.~\ref{nonmarkov.sec} can be mapped onto a system of two coupled
Brownian particles, each equipped with its own heat bath.
In order to further simplify the analysis, we here consider the
overdamped limit $m\dot{v}\to 0$ where the inertia term can be
dropped.
In the overdamped limit, the time evolution equations (\ref{langevin})
simplify to
\begin{eqnarray} \label{overdamped}
  \Gamma_0\dot{x} & = & - \frac{\partial V}{\partial x} - \Gamma(0)x(t)
  +\Gamma(t-t_0)x(t_0) + \eta(t) + h(t) \nonumber\\
  h(t) & = & - \int_{t_0}^t\!ds\, \frac{\partial \Gamma(t-s)}{\partial t}x(s)
  + \zeta(t)
\end{eqnarray}
These equations are studied also in \cite{Cugliandolo_scenario99}.
In \cite{Cugliandolo_scenario99}, the initial time is set
$t_0\to -\infty$, thus
$\Gamma(t-t_0)x(t_0)\to 0$ and the term
depending on the initial condition is absent.

Employing again the exponentially decaying memory function (\ref{memory}),
closed, Markovian time evolution equations for $x$ and $h$ are obtained.
In a generalized notation
\footnote{The original model is recovered if the following identifications
are made: $x_1=x$, $x_2=h$, $U=V+(2\alpha)^{-1}x^2$, $\kappa_2=\alpha$,
$\Gamma_1=\Gamma_0$, $\Gamma_2=\alpha^2\gamma$,
$\nu_1=1/\Gamma_0$, $\nu_2=1/(\alpha^2\gamma)$,
$T_1=T_\mathrm{fast}$, $T_2=T_\mathrm{slow}$.},
the resulting time evolution equations take the form
\begin{eqnarray} \label{twoBrown}
  \dot{x}_1 & = & - \frac{1}{\Gamma_1}\frac{\partial U}{\partial x_1} + \nu_1 x_2 + \eta_1(t)\nonumber\\
  \dot{x}_2 & = & - \frac{\kappa_2}{\Gamma_2}x_2 + \nu_2 x_1 + \eta_2(t)
\end{eqnarray}
where the Gaussian white noise is characterized by $\ave{\eta_i(t)}=0$ and
$\ave{\eta_i(t)\eta_j(s)}=2T_i\Gamma_i^{-1}\delta_{ij}\delta(t-s)$.
In Eq.~(\ref{twoBrown}), we have neglected the transient term
$\Gamma(t-t_0)x(t_0)$, which is justified for times $t$ long enough or
for the initial condition $x(t_0)=0$.
Although Eqs.~(\ref{twoBrown}) are very similar to those studied in
\cite{Cugliandolo_scenario99}, we point out the important difference
that the memory function (\ref{memory}) does not satisfy
$\dot{\Gamma}(0^+)=0$ as is assumed in \cite{Cugliandolo_scenario99}.

Only for the special case
$\Gamma_1\nu_1/T_1=\Gamma_2\nu_2/T_2$
(in terms of the original parameters, this condition reduces to
$T_\mathrm{slow}=T_\mathrm{fast}$)
is the equilibrium distribution function corresponding to (\ref{twoBrown})
a Boltzmann distribution,
$p_\mathrm{eq}(x_1,x_2)\propto \exp{(-U(x_1)/T_1 - \kappa_2x_2^2/T_2 + \nu_1\Gamma_1x_1x_2/T_1)}$.
In that case, the marginal distribution
$f_\mathrm{eq}(x_1)=\int_{-\infty}^\infty\!dx_2\,p_\mathrm{eq}(x_1,x_2)$
is given by
$f_\mathrm{eq}(x_1)\propto \exp{(-U(x_1)/T_1 + \kappa_{21}x_1^2/2T_2)}$
where $\kappa_{21}=(\Gamma_2\nu_2)^2/\kappa_2$.
Thus, the coupling to $x_2$ leads to a shifted equilibrium distribution,
that can be described by a repulsive harmonic potential of strength
$\kappa_{21}$.
In case the potential $U(x_1)$ is itself harmonic,
$U(x_1)=\kappa_1x_1^2/2$,
the shifted equilibrium distribution $f_\mathrm{eq}$ is of the same form
as the uncoupled equilibrium distribution but with a
renormalized spring coefficient.
If instead one insists on the equilibrium form
$f_\mathrm{eq}(x_1)\propto \exp(-U(x_1)/T_\mathrm{eq})$,
an effective temperature 
\footnote{Note, that $T_{\mathrm{eq}}\neq T_1$ if even both baths are at
the same temperature $T_1=T_2$ since the effective temperature
mimics the effect of the coupling on the equilibrium state.
Therefore, quantities that depend only on $x_1$ but not
on $x_2$, show equilibrium expectation values corresponding to
$T_\mathrm{eq}$ rather than $T_1$.} 
is defined by
$T_\mathrm{eq}=T_1/[1 - \Gamma_1\Gamma_2\nu_1\nu_2(\kappa_1\kappa_2)^{-1}]$.
In terms of the original variables, $U$ is harmonic if
the potential $V$ in (\ref{original}) is harmonic,
$V(x)=\kappa_{1,0}x^2/2$. Thus, $\kappa_1=\kappa_{1,0}+\alpha^{-1}$
and the corresponding identification of an effective temperature
equals the bath temperature
$T_\mathrm{eq}=T_1$ as it should, since in this case
the condition $\Gamma_1\nu_1/T_1=\Gamma_2\nu_2/T_2$ reduces to
$T_\mathrm{fast}=T_\mathrm{slow}$.
It is interesting to note, that
the same effective temperature $T_\mathrm{eq} $
is obtained by adiabatic elimination, if $x_2$ is assumed to vary
much faster than $x_1$
\cite{Nieuwenhuizen_efftemp}.
Then, the assumption
$\kappa_1/\Gamma_1\ll\kappa_2/\Gamma_2$
replaces the condition
$\Gamma_1\nu_1/T_1=\Gamma_2\nu_2/T_2$. 

Yet another approach to obtain effective temperatures is to employ
the Quasi-Equilibrium Approximation (QEA)
which is used successfully in many areas of statistical physics
\cite{Zubarevbook,Jou_efftempidealgas}.
Following the standard procedure \cite{Zubarevbook}, extremizing the
entropy functional
$S[p]=-\int\!dx_1dx_2\, p\ln p$ subject to the constraints
of fixed moments
$M_0=\ave{1}$, $M_1=\ave{x_1^2}$, $M_2=\ave{x_2^2}$,
$M_3=\ave{x_1x_2}$, the quasi-equilibrium distribution is found to be
given by
$p^\ast(x_1,x_2)=\exp{(\lambda_0+\lambda_1x_1^2+\lambda_2x_2^2+\lambda_3x_1x_2)}$.
The Lagrange multipliers $\lambda_i$ are chosen such that
the constraints are satisfied identically.
In equilibrium, $p^\ast$ reduces to $p_\mathrm{eq}$.
Out of equilibrium, $p^\ast$ approximates the true non-equilibrium
state by equilibrating all degrees of freedom except for the
macroscopic variables $M_i$.
The quasi-equilibrium entropy $S^\ast$ is defined by
$S^\ast(M)=S[p^\ast]$ and satisfies
$\mathrm{d}S^\ast = \sum_j \lambda_j\mathrm{d}M_j$.
Within the QEA, the temperature is defined as
$T^{-1}_\mathrm{QEA} \equiv \partial S^\ast/\partial U$.
For simplicity, we will assume a harmonic potential,
$V(x)=\kappa_{1,0}x^2/2$.
Then, the effective temperature within the QEA is given by
$T^{-1}_\mathrm{QEA} = (2/\kappa_{1})\lambda_1$.
Thanks to the simplicity of the model, the Lagrange multipliers
can be evaluated explicitly by performing Gaussian integrals.
The resulting expression for $T_\mathrm{QEA}$ is
\begin{equation} \label{T_QEA}
  T_\mathrm{QEA} = \kappa_{1}( \ave{x_1^2} - \ave{x_1x_2}^2/\ave{x_2^2} ).
\end{equation}
In the uncoupled case, $\ave{x_1x_2}=0$,
Eq.~(\ref{T_QEA}) reduces to the equipartition theorem and
$T_\mathrm{QEA}=T_1$.

In the following section, we evaluate and compare different definitions
of effective temperatures derived from the exact solution in case
of harmonic potentials.

\section{Exact solution for harmonic forces} \label{harmonic.sec}
For the harmonic potential $U(x_1)=\kappa_1x_1^2/2$,
$\kappa_1=\kappa_{1,0}+1/\alpha$,
Eqs.~(\ref{twoBrown}) are stochastic differential equations of the
narrow-sense linear type that can be solved exactly by
transformation to normal coordinates.
The solution reads
\begin{equation} \label{xsol}
  x_i(t) = \sum_{\alpha=+,-}b_{i,\alpha}X_\alpha(t)
\end{equation}
where the coefficients $b_{i,\alpha}$ are defined by
$b_{1,+}=-a_-/(a_+-a_-)$, $b_{1,-}=a_+/(a_+-a_-)$, $b_{2,+}=1/(a_+-a_-)$,
$b_{2,-}=-1/(a_+-a_-)$ and
\begin{equation} \label{normalcoords}
  X_\alpha(t) = e^{-(t-t_0)c_\alpha}X_\alpha(t_0)
  + \int_{t_0}^t\!ds\, e^{-(t-s)c_\alpha}\eta_\alpha(s).
\end{equation}
The noise terms appearing in Eq.~(\ref{normalcoords}) are related to those
in (\ref{twoBrown}) by $\eta_\alpha(t)=\eta_1(t)+a_\alpha\eta_2(t)$.
The eigenvalues associated with Eqs.~(\ref{twoBrown}) are given by
\begin{equation}
  c_\pm = \frac{1}{2}\left(
    \frac{\kappa_1}{\Gamma_1} + \frac{\kappa_2}{\Gamma_2} \right)
  \mp \frac{1}{2}\sqrt{D}
\end{equation}
with
$D=(\kappa_1/\Gamma_1-\kappa_2/\Gamma_2)^2+4\nu_1\nu_2$.
The coefficients $a_\pm$ introduced above are defined by
$a_\pm = (\kappa_1/\Gamma_1-c_\pm)/\nu_2$.
Note, that the special case $\nu_2=0$ has to be treated separately.
In that case, the time evolution equation for $x_2$ can be solved
independently of $x_1$, resulting in an additional noise term in
form of an Ornstein-Uhlenbeck process resulting from $x_2$.
In particular, one finds in this case, $c_\pm=\kappa_1/\Gamma_1$ and
$a_\pm=-(\kappa_1/\Gamma_1-\kappa_2/\Gamma_2)^{-1}\nu_1$.
In the uncoupled case, $\nu_1=0$, the eigenvalues reduce to the
individual relaxation frequencies $c_+^0=\kappa_2/\Gamma_2$ and
$c_-^0=\kappa_1/\Gamma_1$.
In the coupled case, $\nu_1\nu_2\neq 0$, stable solutions with
positive eigenvalues $c_\pm$ exist for
$\kappa_1\kappa_2>(\Gamma_1\nu_1)(\Gamma_2\nu_2)$.
For $\nu_1\nu_2<0$ a further condition
$|\kappa_1/\Gamma_1-\kappa_2/\Gamma_2|>2\sqrt{|\nu_1\nu_2|}$
is necessary in order to ensure real values of $c_\pm$.
For the original parameters, $\nu_1=1/\Gamma_0>0$ and the stability
condition reads $\kappa_1\alpha>0$.

In the long-time ('static') limit, we find
from the exact solution (\ref{xsol})
\begin{equation} \label{x2_longtime}
\ave{x_ix_j}_\infty = \sum_{\alpha,\beta}
\frac{2b_{i,\alpha}b_{j,\beta}}{c_\alpha+c_\beta}
\left( \frac{T_1}{\Gamma_1} + a_\alpha a_\beta \frac{T_2}{\Gamma_2}\right)
\end{equation}
In the uncoupled case, $\nu_1=0$, the equipartition theorem
$\tilde{\kappa}_1\ave{x_1^2}_\infty=T_1$ is recovered. Here and in
the following, $\tilde{\kappa}_1=\kappa_{1,0}$ refers to the bare
spring constant of the original model. If, however, one considers
Eqs.~(\ref{twoBrown}) as starting point,
$\tilde{\kappa}_1=\kappa_1$ is interpreted as the bare spring
constant. Note, that in terms of the original model parameters,
the uncoupled case $\nu_1=0$ corresponds to
$\Gamma_0,\alpha\to\infty$. In the general case $\nu_1\neq 0$, one
might assume that some generalization of the equipartition theorem
holds,
\begin{equation} \label{T_static}
  T_\mathrm{static} = \tilde{\kappa}_1\ave{x_1^2}_\infty
\end{equation}
with an effective ('static') temperature
$T_\mathrm{static}$.
Note, however, that such an effective temperature results from an
underlying canonical distribution function
only for the special choice
$\Gamma_1\nu_1/T_1=\Gamma_2\nu_2/T_2$,
as was demonstrated in Sec.~\ref{overdamped.sec}.
For this choice of parameters,
$T_\mathrm{static}=T_\mathrm{slow}$.
For a general choice of parameters, $T_\mathrm{static}$
denotes the effective temperature deduced from the equipartition
theorem if we would not be aware of the coupling of $x_1$ to $x_2$.
A comparison of these and other definitions of effective temperatures
is provided later.

Next, we consider possible definitions of effective temperatures
from time-dependent \FDR.
The response of the system (\ref{twoBrown}) to a time-dependent
perturbation $f(t)$ is measured by the response functions
\begin{equation}
  R_{i,1}(t,t') \equiv \left. \frac{\delta \ave{x_j(t)}_f}{\delta f(t')}
  \right|_{f=0}
\end{equation}
Due to the simplicity of the model, the response functions are  also
obtained analytically,
\begin{equation} \label{response}
  R_{i,1}(t,t') =
  \Gamma_1^{-1}\sum_{\alpha=+,-} b_{i,\alpha}e^{-(t-t')c_\alpha}\Theta(t-t')
\end{equation}
where $\Theta(t)$ denotes the unit step function.
The response functions are causal, $R(t,t')=0$ for $t<t'$, and
time-translational invariant, $R_{i,1}(t,t')=R_{i,1}(t-t')$,
as they should.
The correlation functions are defined by
\begin{equation} \label{correl}
  C_{i,j}(t,t') \equiv \ave{x_i(t)x_j(t')} - \ave{x_i(t)}\ave{x_j(t')}
\end{equation}
Inserting the exact solution, Eqs.~(\ref{xsol}), (\ref{normalcoords}),
into (\ref{correl}), one obtains
\begin{equation}   \label{C_t}
  C_{i,j}(t,t') = \sum_{\alpha,\beta}
  \frac{2b_{i,\alpha}b_{j,\beta}}{c_\alpha+c_\beta}
  \left( \frac{T_1}{\Gamma_1}
    + a_\alpha a_\beta \frac{T_2}{\Gamma_2} \right)
  \left[ e^{-(t-t')c_\alpha} - e^{-(t-t_0)c_\alpha}e^{-(t'-t_0)c_\beta} \right]
\end{equation}
where $t>t'$ has been assumed without loss of generality.
Again, due to time-translational invariance
$C_{i,j}(t,t')=C_{i,j}(t-t')$ holds.
For the special choice of parameters corresponding to the
original model, Eqs.~(\ref{response}) and
(\ref{C_t}) are equivalent to the equations given in
\cite{Cugliandolo_FluktTheorem}.
Note, that the long time limit of the equal time correlation function
reduces to $\lim_{t\to\infty}C_{i,j}(t,t)=\ave{x_ix_j}_\infty$.
From Eq.~(\ref{C_t}) we observe that the decay of the correlation function
is determined by the two eigenvalues $c_\pm$. This can be interpreted
as a two-step process, where the initial, fast decay is described
by the larger eigenvalue $c_-$, while the smaller eigenvalue $c_+$
governs the long time decay. Such an interpretation is particularly
relevant if the two eigenvalues are well-separated.

The fluctuation-dissipation relation (\FDR) is most conveniently expressed
in terms of the integrated response function, defined by
$\chi_{i,j}(t)=\int_{-\infty}^t\!ds\,R_{i,j}(s)$.
Integrating Eq.~(\ref{response}) over time differences one obtains
\begin{equation} \label{chi_t}
  \chi_{i,1}(t) = \Gamma_1^{-1} \sum_{\alpha=+,-}
  \frac{b_{i,\alpha}}{c_\alpha}\left(
  1 - e^{-tc_\alpha} \right).
\end{equation}

In equilibrium, one can prove under quite general assumptions the
validity of the \FDT,
\begin{equation} \label{FDT}
  \tilde{\chi}_{i,j}(t) = \frac{1}{T}[1 - \tilde{C}_{i,j}(t)]
\end{equation}
where $\tilde{\chi}_{i,j}(t)\equiv\chi_{i,j}(t)/C_{i,j}(t=0)$  and
$\tilde{C}_{i,j}(t)\equiv C_{i,j}(t)/C_{i,j}(t=0)$.
In the present case, the \FDT holds identically in the uncoupled case
$\nu_1=0$.
For $\nu_1\neq 0$, however, the \FDT is violated for the system studied
here, as can be seen directly by comparing
Eqs.~(\ref{C_t}) and (\ref{chi_t}).
It was proposed in \cite{Cugliandolo_effTemp97},
that a meaningful effective temperature can be defined from the
\FDR by
\begin{equation} \label{Teff_FDT}
  -T_\mathrm{eff}^{-1} = \frac{\mathrm{d}\chi_{i,j}}{\mathrm{d}C_{i,j}}.
\end{equation}
For short times, the effective temperature defined from
Eq.~(\ref{Teff_FDT}) coincides with $T_1$, as is readily shown by
expanding the exponentials in Eqs.~(\ref{C_t}) and (\ref{chi_t})
to first order in $tc_\alpha$. For long times, the time dependence
of both the correlation and response function is dominated by the
smallest eigenvalue. In this regime, the effective temperature
defined from Eq.~(\ref{Teff_FDT}) is given by
\begin{equation} \label{Teff_long}
  T_\mathrm{eff}^\mathrm{long} = \left(
    \frac{\kappa_1}{\Gamma_1}+\frac{\kappa_2}{\Gamma_2}\right)^{-1}
  \left(T_1 \left[c_+ + \frac{\kappa_2}{\Gamma_2}\right]
  + T_2\frac{\Gamma_1\nu_1}{\Gamma_2\nu_2}\left[c_- - \frac{\kappa_2}{\Gamma_2}\right] \right)
\end{equation}

Finally, we introduce also the effective temperature
$T_\infty^{-1}=\chi_{1,1}(t\to\infty)/C_{1,1}(0)$,
$C_{1,1}(0)=\ave{x_1^2}_\infty$,
introduced in \cite{OHern2004}.
This temperature, defined from the 'static' limit of the \FDR,
was found to be useful when the definition (\ref{Teff_long})
was problematic \cite{OHern2004}.
From the long time limit of (\ref{chi_t}) we find
\begin{equation}  \label{Too}
  T_\infty = T_\mathrm{static}(\kappa_1/\tilde{\kappa}_1) \left[
  1 - \Gamma_1\Gamma_2\nu_1\nu_2(\kappa_1\kappa_2)^{-1}\right]
\end{equation}
where $T_\mathrm{static}$ is the static, effective temperature
defined in (\ref{T_static}).
Note, that $T_\infty$ is well-defined since we assumed
$\kappa_1\kappa_2>(\Gamma_1\nu_1)(\Gamma_2\nu_2)$ above.
Within the original model, $T_\infty=T_\mathrm{static}$ 
holds which is directly verified inserting the original 
model parameters.

\begin{table}
\caption{Overview of different definitions of effective
temperatures.}
\begin{center}
\begin{tabular}{|l|l|c|}
\br
symbol & effective temperature from & defined in \\
\mr
$T_\mathrm{QEA}$ & quasi-equilibrium approximation & (\ref{T_QEA})\\
\hline
$T_\mathrm{static}$ & equipartition theorem & (\ref{T_static})\\
\hline
$T_\mathrm{eff}$ & \FDR & (\ref{Teff_FDT})\\
\hline
$T_\mathrm{eff}^\mathrm{long}$ & long-time limit of the \FDR & (\ref{Teff_long})\\
\hline
$T_\infty$ & 'static' limit of the \FDR & (\ref{Too})\\
\br
\end{tabular}
\end{center}
\label{Teffs.table}
\end{table}

Table \ref{Teffs.table} summarizes the different definitions of
effective temperature considered here.

\begin{figure}
\begin{minipage}{18pc}
\includegraphics[width=18pc]{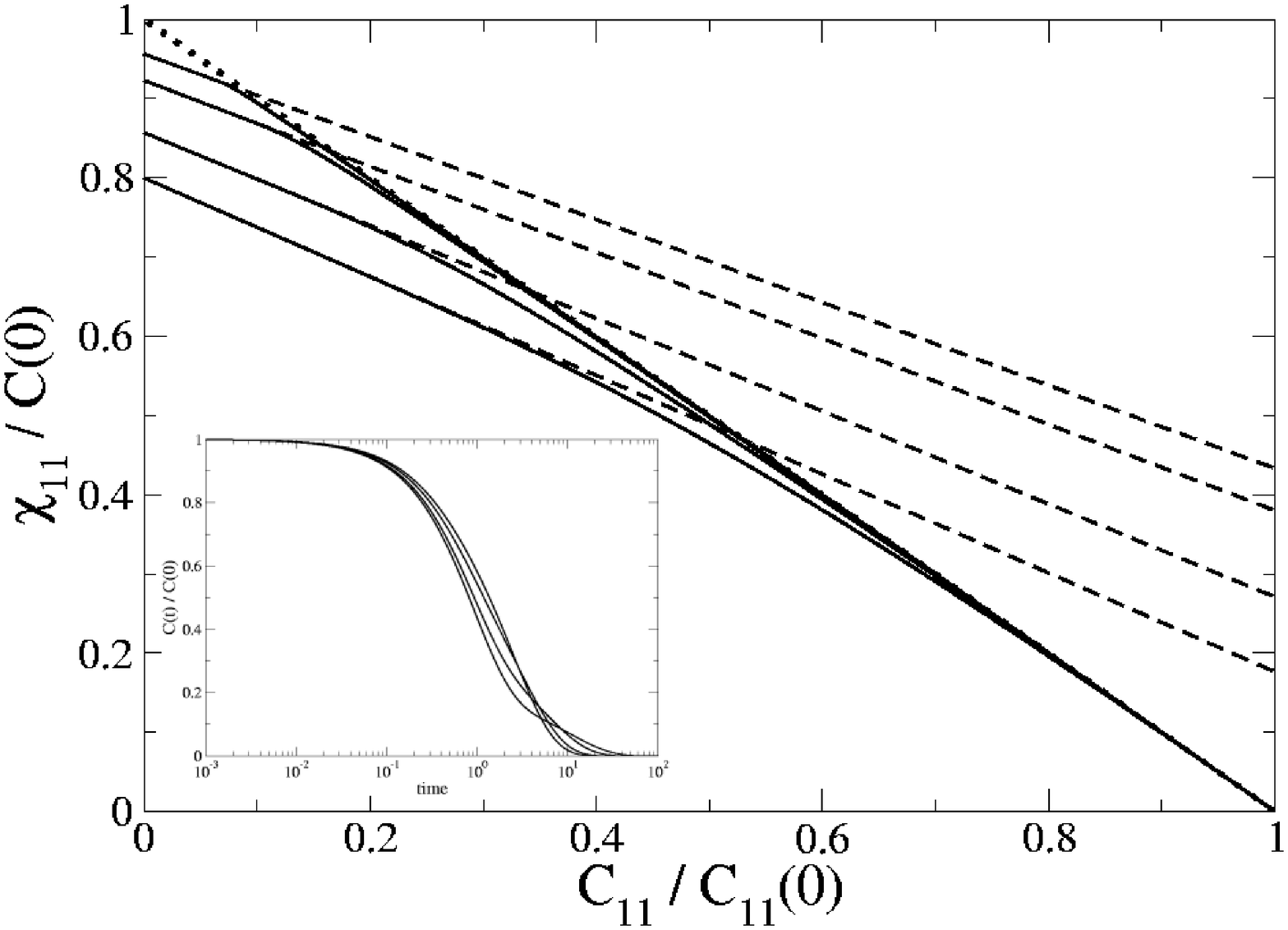}
\end{minipage}\hspace{2pc}%
\begin{minipage}{18pc}
\includegraphics[width=18pc]{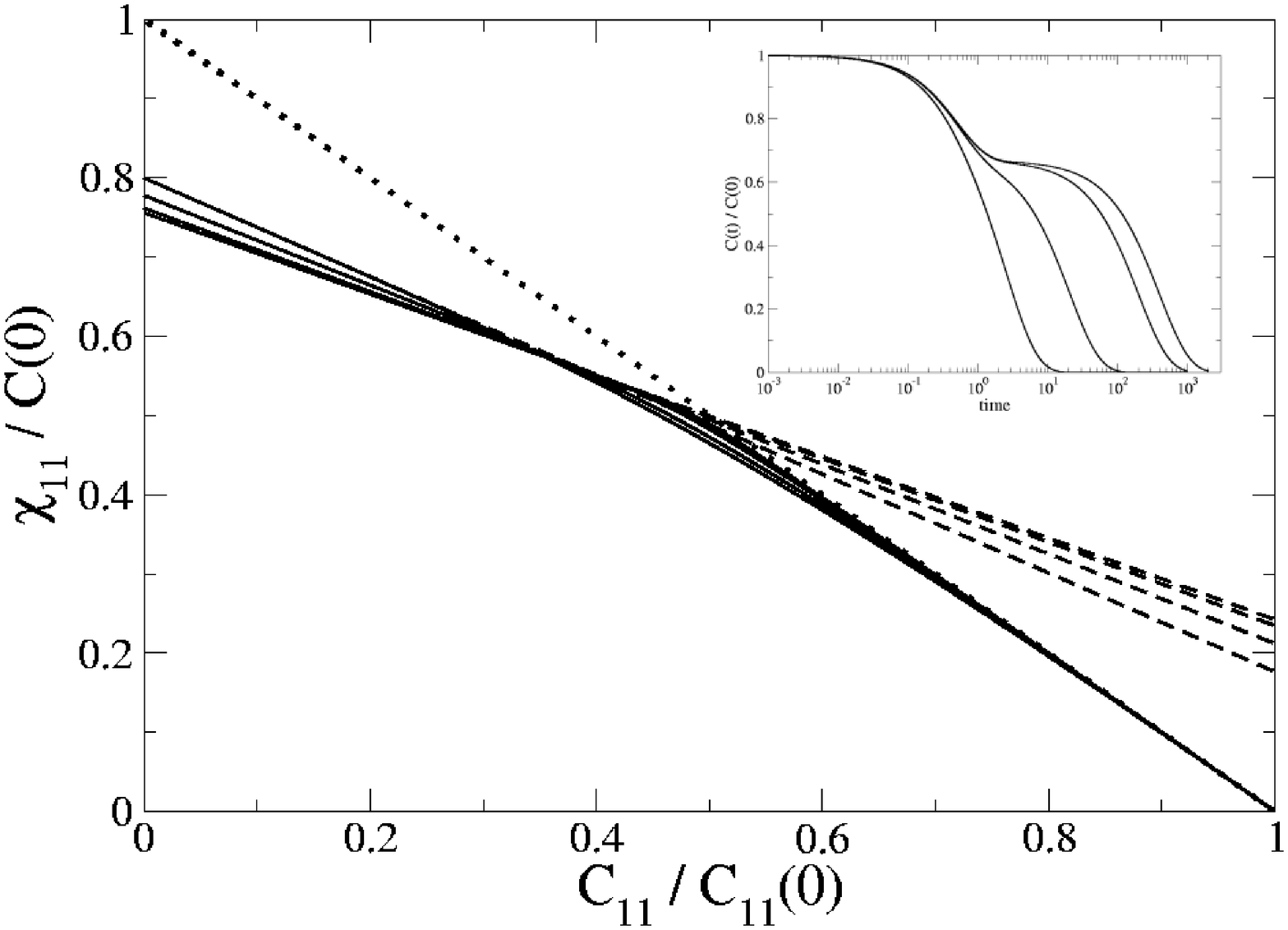}
\end{minipage}
\caption{The integrated response function (\ref{chi_t}) is plotted
versus the correlation function (\ref{C_t}).
Dashed lines have inverse slopes calculated from
Eq.~(\ref{Teff_long}).
The inset shows the time evolution of the normalized correlation
function for the same parameters.
The parameters have been chosen as
$\alpha=\gamma=\Gamma_0=\kappa_{1,0}=\kappa_2=1$, $T_\mathrm{fast}=1$, and
$T_\mathrm{slow}=2$ if not stated otherwise.
In the left figure, the values of $\alpha$ from bottom to top are
$\alpha=1, 2, 5, 10$, while $\gamma$ varies from top to bottom
on the far left of the right figure as $\gamma = 1, 2, 10, 100$.}
\label{chi_vs_C.fig}
\end{figure}

In Fig.~\ref{chi_vs_C.fig}, the integrated response
$\chi_{1,1}(t)$ is plotted in a parametric plot,  versus the
correlation function $C_{1,1}(t)$. Parameters are chosen as
$\gamma=\Gamma_0=T_\mathrm{fast}=1$,
$\kappa_{1,0}=T_\mathrm{slow}=2$. The values of $\alpha$ are
varied between $1\leq \alpha\leq 10$. From Fig.~\ref{chi_vs_C.fig}
one observes a cross-over from the short time regime with
$T_\mathrm{eff}=T_1$ to the long time regime with $T_\mathrm{eff}$
given by (\ref{Teff_long}). This scenario has been found also for
spin glass systems \cite{Cugliandolo_effTemp97} and in molecular
dynamics simulations of sheared model-glasses
\cite{Barrat_FDT2002}. For increasing coefficients $\alpha$ or
$\kappa_{1,0}$, $T_\mathrm{eff}$ increases and the cross-over is
shifted to later times. For increasing $\gamma$, $T_\mathrm{eff}$
increases as well, however, the cross-over point is not shifted.

\begin{figure}
\centering
\includegraphics[width=10cm]{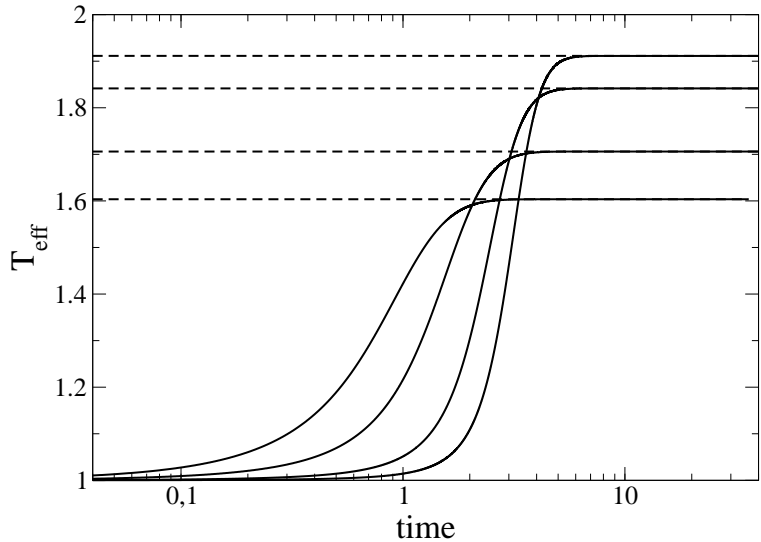}
\caption{The time evolution of the effective temperature defined from
the time-dependent \FDR, Eq.~(\ref{Teff_FDT}), is shown
on a logarithmic time scale. The same values of the
parameters as in Fig.~\ref{chi_vs_C.fig} (left panel) are used.
Dashed lines are the long-time effective temperatures calculated from
Eq.~(\ref{Teff_long}).}
\label{Teff_t.fig}
\end{figure}

The time evolution of the effective temperature (\ref{Teff_FDT})
is shown in Fig.~\ref{Teff_t.fig}, where $T_\mathrm{eff}$ is
calculated with the help of the time derivatives of
Eqs.~(\ref{C_t}) and (\ref{chi_t}). Equilibrium initial conditions for the
uncoupled system were chosen.
Figure \ref{Teff_t.fig} shows the approach of the effective
temperature to its long time asymptotic value (\ref{Teff_FDT}).
For the present choice of parameters, the asymptotic value is
approached for times $t\gtrsim 1$, while $T_\mathrm{eff}$
remains close to the bath temperature $T_1$ for short times.

\begin{figure}
\centering
\includegraphics[width=12cm]{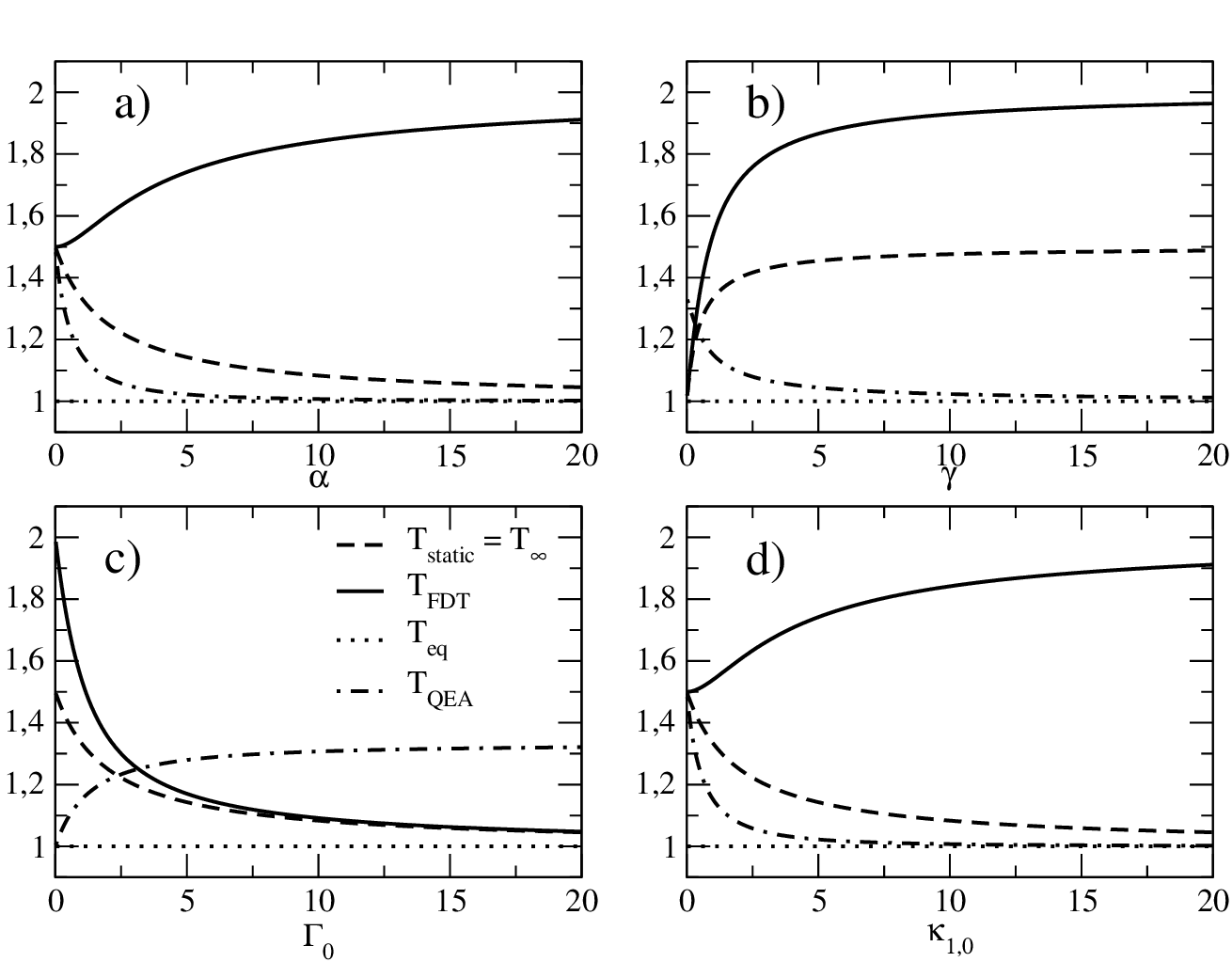}
\caption{Different definitions of effective temperatures are
shown as a function of model parameters
$\alpha, \gamma, \Gamma_0$, and $\kappa_{1,0}$.
The remaining parameters are chosen as in Fig.~\ref{chi_vs_C.fig},
with $\alpha=1$ and $\gamma=1$.}
\label{Teff_static_g1.fig}
\end{figure}

\begin{figure}[b]
\centering
\includegraphics[width=12cm]{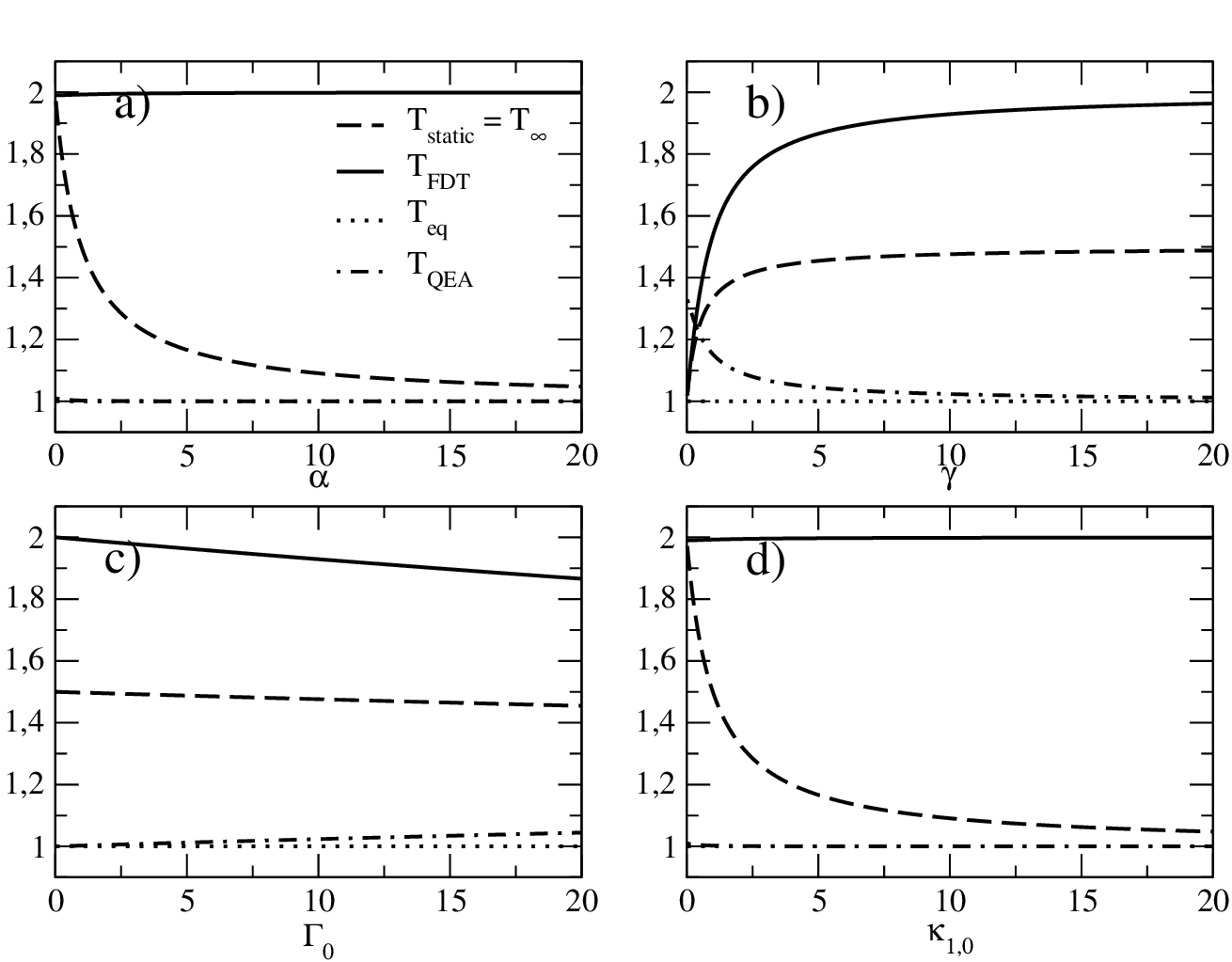}
\caption{Different definitions of effective temperatures are
shown as a function of model parameters
$\alpha, \gamma, \Gamma_0$, and $\kappa_{1,0}$.
The remaining parameters are chosen as in Fig.~\ref{chi_vs_C.fig},
with $\alpha=1$ and $\gamma=100$.}
\label{Teff_static_g100.fig}
\end{figure}

Figures \ref{Teff_static_g1.fig}, \ref{Teff_static_g100.fig} 
show a comparison of different definitions
of effective temperatures for several model parameters.
We observe that for increasing $\alpha$, or $\gamma$ or $\kappa_{1,0}$,
the static temperatures all approach the temperature $T_\mathrm{fast}$ of
the fast
bath, while the effective temperature from the long-time asymptotics of the
\FDR approaches the temperature $T_\mathrm{slow}$ of the slow bath.
Only in case of very strong friction coefficient $\Gamma_0$ associated
with the
fast bath, does $T_\mathrm{eff}$ fail to approach $T_\mathrm{slow}$.

\section{Conclusion} \label{concl.sec}

The concept of effective temperatures has been studied within an
exactly solvable model of non-Markovian dynamics for a harmonic
oscillator coupled to two heat baths held at different
temperatures. For parameter ranges ($\alpha\gg 1$, or $\gamma\gg
1$, or $\kappa_{1,0}\gg 1$) corresponding to a separation of time
scales of the fast and slow bath, it has been found that the
effective temperature $T_\mathrm{eff}^\mathrm{long}$ defined from
the long-time limit of the fluctuation-dissipation relation (\FDR)
indeed approaches the temperature of the slow heat bath. This
result is therefore consistent with the proposition
\cite{Cugliandolo_effTemp97} of $T_\mathrm{eff}^\mathrm{long}$ as
a useful definition of effective temperatures out of equilibrium. 
For intermediate values of the parameters, when the separation of
time scales is not achieved, the effective temperature
$T_\mathrm{eff}^\mathrm{long}$ is intermediate between the
temperature of the slow and fast heat bath. Only in the unlikely
case of a friction coefficient $\Gamma_0$ associated with the fast
bath, larger than the friction $\gamma$  associated with the slow
bath, $T_\mathrm{eff}^\mathrm{long}$ approaches the temperature of
the fast bath, while the quasi-equilibrium temperature is somewhat
higher.

In comparison, other,  static effective temperature definitions
yield results that are much more difficult to rationalize. In most
cases, the temperature of the fast bath is obtained from these
definitions.  However, in some cases,  values intermediate between
$T_\mathrm{slow}$ and $T_\mathrm{fast}$
are obtained even when time scales are well separated (see e.g.
figures 3 and 4, panels b and c, for $T_\mathrm{static}$ or 
$T_\mathrm{QEA}$).

We finally   mention that a breakdown of the effective temperature
from the long-time limit of the \FDR can be observed in the model
(\ref{twoBrown}) for particular choices of the parameters that
are, however, inconsistent with the original model
\ref{overdamped}. For couplings $\nu_1<0$, $\nu_2>0$,
$\Gamma_1<\Gamma_2$ and $T_1>T_2$, there are parameter ranges
where an overshoot in the response function $\chi$ is observed
together with a corresponding undershoot in the correlation
function $C$. There, the parametric plot $\chi$ versus $C$
continues to negative values of $C$ and $\chi(C)$ becomes
multi-valued. Approaching this regime, the plot $\chi$ versus $C$
flattens more and more, which makes the application of
(\ref{Teff_FDT}) difficult. A similar situation was observed for
some observables in molecular dynamics simulations
\cite{OHern2004}.

The present study confirms and extends previous results
\cite{Jou_efftempidealgas} that several, non-equivalent
effective temperatures can be defined. In the model
system considered here, which could be taken as a model for an
internal degree of freedom in a slowly driven system, the
definition that appears to have the most sensible behavior is
associated with the long time limit of the fluctuation dissipation
ratio.

\ack

P.I. acknowledges financial support from the Alexander von Humboldt
foundation.

\section*{References}

\newpage

\newpage

\begin{thebibliography}{10}

\bibitem{Cugliandolo_effTemp97}
L.~Cugliandolo, J.~Kurchan, and L.~Peliti.
\newblock Energy flow, partial equilibrium, and effective temperatures in
  systems with slow dynamics.
\newblock {\em Phys. Rev. E}, 55(4):3898, 1997.

\bibitem{Nieuwenhuizen_efftemp}
T.~M. Nieuwenhuizen.
\newblock Thermodynamics of the glassy state: effective temperature as
  additional system parameter.
\newblock {\em Phys. Rev. Lett.}, 80:5580--5583, 1998.

\bibitem{KobBarrat2000}
W.~Kob and J.-L. Barrat.
\newblock Fluctuations, response and aging in a simple glass-forming liquid out
  of equilibrium.
\newblock {\em Eur. Phys. J. B}, 13:319--333, 2000.

\bibitem{Barrat_FDT2002}
L.~Berthier and J.-L. Barrat.
\newblock Nonequilibrium dynamics and fluctuation-dissipation relation in a
  sheared fluid.
\newblock {\em J. Chem. Phys.}, 116(4):6228--6242, 2002.

\bibitem{Jou_Review_nonequiTemp2003}
J.~Casas-V\'azquez and D.~Jou.
\newblock Temperature in non-equilibrium states: a review of open problems and
  current proposals.
\newblock {\em Rep. Prog. Phys.}, 66:1937--2023, 2003.

\bibitem{OHern2004}
C.~S. O'Hern, A.~J. Liu, and S.~R. Nagel.
\newblock Effective temperatures in driven systems: Static versus
  time-dependent relations.
\newblock {\em Phys. Rev. Lett.}, 93:165702, 2004.

\bibitem{CristantiRitort_efftempreview}
A.~Crisanti and F.~Ritort.
\newblock Violation of the fluctuation-dissipation theorem in glassy systems:
  basic notions and the numerical evidences.
\newblock {\em J. Phys. A, Math. Gen.}, 36:R181--R290, 2003.

\bibitem{Herrison2002}
D.~H\'errison and M.~Ocio.
\newblock Fluctuation-dissipation ratio of a spin glass in the aging regime.
\newblock {\em Phys. Rev. Lett.}, 88(2):257202, 2002.

\bibitem{Makse2005}
C.~Song, P.~Wang, and H.~A. Makse.
\newblock Experimental measurement of an effective temperature for jammed
  granular materials.
\newblock {\em Proc.~Nat.~Acad.~Sci.}, 102(7):2299--2304, 2005.

\bibitem{Bonn2005}
S.~Jabbari-Farouji, D.~Mizuno, M.~Atakhorrami, F.~C. MacKintosh, C.~F. Schmidt,
  E.~Eiser, G.~H. Wegdam, and D.~Bonn.
\newblock Fluctuation-dissipation theorem in an aging colloidal glass.
\newblock {\em cond-mat/0511311}, 2005.

\bibitem{Strachan2005}
D.~R. Strachan, G.~C. Kalur, and S.~R. Raghavan.
\newblock Two distinct time-scale regimes of the effective temperature for an
  aging colloidal glass.
\newblock {\em cond-mat/0510742}, 2005.

\bibitem{AbouGallet2004}
B.~Abou and F.~Gallet.
\newblock Probing a nonequilibrium einstein relation in an aging colloidal
  glass.
\newblock {\em Phys. Rev. Lett.}, 93:160603, 2004.

\bibitem{Bellon2002}
L.~Bellon and S.~Ciliberto.
\newblock Experimental study of the fluctuation-dissipation-relation during an
  aging process.
\newblock {\em Physica D}, 168:325, 2002.

\bibitem{Cugliandolo_PRL93}
L.~Cugliandolo and J.~Kurchan.
\newblock Analytical solution of the off-equilibrium dynamics of a long range
  spin-glass model.
\newblock {\em Phys. Rev. Lett.}, 71:173, 1993.

\bibitem{Gambassi}
A.~Gambassi.
\newblock Slow dynamics at critical points: the field-theoretical perspective.
\newblock {\em J. Phys. Conf. Series}, 2006.
\newblock this volume.

\bibitem{Schehr}
G.~Schehr and R.~Paul.
\newblock Non-equilibrium critical dynamics in disordered ferromagnets.
\newblock {\em J. Phys. Conf. Series}, 2006.
\newblock this volume.

\bibitem{Sollich_condmat01}
P.~Sollich, S.~Fielding, and P.~Mayer.
\newblock Fluctuation-dissipation relations and effective temperatures in
  simple non-mean field systems.
\newblock 14:1683--1696, 2002.

\bibitem{Cugliandolo_FluktTheorem}
F.~Zamponi, F.~Bonetto, L.~F. Cugliandolo, and J.~Kurchan.
\newblock Fluctuation theorem for non-equilibrium relaxational systems driven
  by external forces.
\newblock {\em cond-mat/0504750}, 2005.

\bibitem{Cugliandolo_scenario99}
L.~Cugliandolo and J.~Kurchan.
\newblock A scenario for the dynamics in the small entropy production limit.
\newblock {\em J.~Phys.~Soc.~Japan}, 69:247, 2000.

\bibitem{Nieuwenhuizen_steadystatedifftimescales}
A.~E. Allahverdyan and T.~M. Nieuwenhuizen.
\newblock Steady adiabatic state: Its thermodynamics, entropy production,
  energy dissipation, and violation of {O}nsager relations.
\newblock {\em Phys. Rev. E}, 62:000845, 2000.

\bibitem{Barrat_nonmarkov90}
J.-L. Barrat.
\newblock Numerical simulation of {B}rownian motion with frequency-dependent
  friction.
\newblock {\em Chem. Phys. Lett.}, 165(6):551--553, 1990.

\bibitem{Zubarevbook}
D.~Zubarev, V.~Morozov, and G.~R\"opke.
\newblock {\em Statistical Mechanics of Nonequilibrium Processes}, volume~1.
\newblock Akademie Verlag, Berlin, 1997.

\bibitem{Jou_efftempidealgas}
D.~Jou, M.~Criado-Sancho, and J.~Casas-V\'azquez.
\newblock Nonequilibrium temperature and fluctuation-dissipation temperature in
  flowing gases.
\newblock {\em Physica A}, 358:49--57, 2005.

\end{thebibliography}
\end{document}